\documentclass[11pt, a4paper]{article}
\pdfoutput=1
\usepackage{jheparxiv}
\usepackage[latin1]{inputenc}
\usepackage{amsmath}
\usepackage{amsfonts}
\usepackage{amssymb}
\usepackage{latexsym}
\usepackage{mathrsfs}
\usepackage{graphicx}
\usepackage{color}
\usepackage{slashed}
\usepackage{twistor}

\renewcommand{\d}{\mathrm{d}}

\subheader{\hfill \texttt{Scholarpedia 9(10):31791}}

\title{The Kerr-Newman metric: A Review}

\author[a]{Tim Adamo}
\author[b]{and E. T. Newman}

\affiliation[a]{Department of Applied Mathematics \& Theoretical Physics, University of Cambridge \\
        Wilberforce Road, Cambridge CB3 0WA, United Kingdom}
\affiliation[b]{Department of Physics \& Astronomy, University of Pittsburgh \\
        100 Allen Hall, 3941 O'Hara St, Pittsburgh PA 15260, USA}

\emailAdd{t.adamo@damtp.cam.ac.uk}
\emailAdd{newman@pitt.edu}

\abstract{The Kerr-Newman metric describes a very special rotating, charged mass and is the most general of the asymptotically flat stationary `black hole' solutions to the Einstein-Maxwell equations of general relativity.  We review the derivation of this metric from the Reissner-Nordstr\"om solution by means of a complex transformation algorithm and provide a brief overview of its basic geometric properties.  We also include some discussion of interpretive issues, related metrics, and higher-dimensional analogues.}

\begin{document}

\maketitle

\section{Background \& Introduction}

In four space-time dimensions, the \emph{no hair} theorem dictates that all black hole solutions to the Einstein-Maxwell equations are uniquely characterized by three numbers: mass, electric charge, and angular momentum.  In the static case, where the angular momentum vanishes and there is a spherical symmetry, we have the well-known Schwarzschild and Reissner-Nordstr\"om metrics depending on whether there is electric charge.  The spinning generalization of the Schwarzschild solution is the Kerr metric \cite{Kerr:1963ud}, while the charged spinning solution, or \emph{Kerr-Newman} metric \cite{Newman:1965my}, is both the spinning generalization of Reissner-Nordstr\"om and the electrically charged version of the Kerr metric.  These four metrics are often referred to as the `black hole' solutions of general relativity.

These solutions strongly suggest that charged and rotating bodies can undergo gravitational collapse to form black holes just as in the uncharged, static case of the Schwarzschild metric.  Of course, the greater relevance of these solutions for astrophysical applications is debatable: the electric charge of physical black holes, stars, or planets is screened from distant observers by the accretion of matter with counterbalancing charge, and the assumption of axial symmetry forces the magnetic dipole moment of the solution to align with the rotation axis.  


Nevertheless, the Kerr-Newman metric is the most general static/stationary black hole solution to the Einstein-Maxwell equations.  As such, it is clearly of great importance for theoretical considerations within the mathematical framework of general relativity and beyond.  Furthermore, understanding this solution also provides valuable insights into the other black hole solutions, in particular the Kerr metric.  

This article is not meant to serve as a comprehensive review of all black hole solutions, the formalisms used to derive them, or even all aspects of the Kerr-Newman metric itself.  Rather, we aim to provide a coherent account of how this metric can be derived by a complex transformation algorithm, as well as a discussion of its relevant features and some related topics.  A familiarity with basic concepts in general relativity will be assumed, and we often invoke the spin-coefficient, or Newman-Penrose, formalism for general relativity \cite{Newman:1961qr}, which expresses the field equations in terms of the complex spin connection rather than the metric.  Reviews on this topic can be found in the associated Scholarpedia article \cite{Newman:2009}, or any number of other sources (c.f., \cite{Newman:1981fn, Penrose:1984, Penrose:1986ca, Adamo:2009vu}).  Readers interested in more expansive treatments may find the textbooks \cite{Wiltshire:2009, Stephani:2003} useful; the former reviews many aspects of the Kerr solution, while the latter contains a particularly detailed account of the Petrov classification for algebraically special space-times, the Kerr-Schild formalism, and general solution generating techniques. A concise account of the Kerr metric can also be found in \cite{Teukolsky:2014vca}.

\medskip

The Kerr metric was originally derived by imposing the condition of algebraic speciality on top of stationarity and axial symmetry of the Einstein equations.\footnote{Kerr's original ansatz was for a single repeated null direction, or Petrov type II, although the Kerr metric is actually of type D (c.f., \cite{Kerr:2007dk}).}  However, it was soon noticed that the Kerr solution could also be found by applying an unusual complex coordinate transformation to the Schwarzschild metric \cite{Newman:1965tw}.  This was motivated by an observation associated with Petrov type D algebraically special metrics.  One of the degenerate principal null vectors (DPNVs) of the Schwarzschild metric generates a congruence whose complex divergence (given by the single spin-coefficient, $\rho$) is real and of the form $\rho=-r^{-1}$.  For the Kerr metric, the same spin-coefficient is complex, given by $\rho=-(r-ia\cos\theta)^{-1}$.  For algebraically special metrics, $\rho$ plays a dominant role in the radial behavior of almost all other variables.

Hence, it seemed likely that a transformation $r\rightarrow r+ia\cos\theta$ could take the Schwarz--schild to the Kerr metric; with a few subtleties, this was seen to be true \cite{Newman:1965tw}.  Since the Reissner-Nordstr\"om metric also has a DPNV with complex divergence $\rho=-r^{-1}$, it seemed natural to perform the same complex transformation in that setting, and this led to the charged spinning metric, or Kerr-Newman metric \cite{Newman:1965my}.  In both rotating cases, the transformation sends the DPNV congruence from a gradient to a twisting congruence.  Of course, the associated Maxwell field is required to obtain a full solution of the Einstein-Maxwell equations; in the original work this derivation was non-trivial.

Needless to say, there ought to be a deeper reason for why the complex transformation procedure works.  A partial answer is given by the fact that all of the black hole solutions are examples of \emph{Kerr-Schild} metrics \cite{Kerr:1965}, and such metrics can be endowed with electric charge without modifying their associated null congruence \cite{Debney:1969zz}.  Hence, Schwarzschild can be associated to Reissner-Nordstr\"om and likewise Kerr to Kerr-Newman; once the link between Schwarzschild and Kerr is established, it is clear that it can be extended to the charged setting, although doing so in practice is non-trivial. 

Although the efficacy of the complex transformation has often been considered a `trick' or an accident, it can also be viewed as one early example of using complexified space-time to derive potentially significant results.  The general concept of working with a complexified space-time and imposing reality conditions at a later stage can be seen in a wide variety of works, including (but not limited to) twistor theory \cite{Penrose:1967wn, Penrose:1972ia}, the study of asymptotically flat space-times \cite{Newman:1976gc, Hansen:1978jz, Adamo:2009vu}, or even scattering amplitudes in gauge theory and gravity (c.f., \cite{Elvang:2013cua}).  



\medskip

The history of the Kerr-Newman metric's derivation is itself an interesting story, and is enmeshed with the discovery of two other well-known exact solutions to the Einstein field equations: the Kerr and Newman-Unti-Tamburino metrics (see \cite{Kerr:2007dk} for another account of this story).

The construction of the spin-coefficient (or Newman-Penrose) formalism of general relativity proved highly useful from the perspective of investigating algebraically special metrics \cite{Newman:1961qr}.  In the spin-coefficient language, the standard Einstein equations for the metric are replaced by a much larger set of equations for many (complex and real) extra variables; this is tantamount to working with the complex spin-frame of space-time instead of the metric.  The advantage of the increased number of variables is that all the field equations are first-order, and restrictive conditions can be imposed--before any calculations are begun--to limit the search for solutions of a given class.  In particular, the requirement that the space-time be algebraically special is easily implemented in this way.  This allowed for simple re-derivations of the Goldberg-Sachs theorem \cite{Goldberg:1962} and Robinson-Trautman metrics \cite{Robinson:1962zz}, in addition to novel solutions (e.g., geodesic rays \cite{Newman:1962}, Demia\'nski-Newman \cite{Demianski:1966}).

In 1962, Newman, Tamburino, and Unti submitted a paper to \textit{Journal of Mathematical Physics} which exploited the spin-coefficient formalism to derive two results.  One was the discovery of a new stationary, axisymmetric, homogeneous space-time (shown to be a coordinate extension of an earlier metric found by Taub \cite{Taub:1950ez}) which is now referred to as the Taub-NUT metric.  The second was a theorem claiming that a particular class of type II twisting metrics did \emph{not} exist.  This claim rested on a certain constant of integration, $a$, being forced to vanish by a relation akin to $a=-a$.  Unfortunately, this relation actually followed from a sign error in one of the many field equations of the spin-coefficient formalism (c.f., \cite{Newman:1963er}).

This draft was sent for peer-review to Alfred Schild, who in turn passed it on to Kerr.  Kerr detected this error and, with Newman, eventually identified its source in the spin-coefficient equations; with the corrected sign, there was no constraint that $a$ should vanish, and this would eventually become the Kerr parameter.  The manuscript was revised, and only the NUT metric (along with acknowledgements thanking Kerr and Alan Thompson for catching this serious error) appeared in published form \cite{Newman:1963yy}.  Simultaneously, Kerr used his knowledge of the error in the prior draft to work out the correct twisting metric, which depended on two parameters (a mass $M$ and the non-vanishing $a$) \cite{Kerr:1963ud}.

Using the corrected version of the calculation that had been attempted in the draft as well as Kerr's form of the metric, one could now see (almost by inspection alone) a way to pass from the Schwarzschild to Kerr metric by a complex coordinate transformation \cite{Newman:1965tw}.  This in turn could be extended by an easy `guess' to provide an algorithm for going from the Reissner-Nordstr\"om metric to a new, then unknown, metric for a charged spinning particle \cite{Newman:1965my}.

\medskip

We begin in Section \ref{Derivation} by deriving the Kerr-Newman metric from the Reissner- Nordstr\"om solution using the complex coordinate transformation of \cite{Newman:1965tw}.  After first applying this algorithm to obtain the metric tensor, we then use a recent extension of the original algorithm due to Keane \cite{Keane:2014} which, in addition to the metric, also produces the appropriate Maxwell field and Weyl tensor components in the spin-coefficient formalism.\footnote{Other formulations of the algorithm exist which can be extended to the Maxwell field, (c.f., \cite{Erbin:2014aya}).}  The veracity of the algorithm is confirmed by direct substitution into the spin-coefficient equations.  Some other aspects and extensions are discussed.

Section \ref{Properties} then outlines some of the basic features of the Kerr-Newman solution, including its expression in different coordinate systems (Boyer-Lindquist, Kerr-Schild), singularity structure, event horizons, and ergosphere. We conclude with Section \ref{Advances}, which provides a brief (and by no means exhaustive) outline of progress and results related to the Kerr-Newman metric which have appeared since its discovery in 1965.  These include various interpretive issues, related solutions, and generalizations in higher dimensions.

Throughout, space-time indices will be denoted by Greek letters in the middle of the alphabet (e.g., $\mu, \nu,\ldots$), and the metric has signature $(+,-,-,-)$.

        
\section{Derivation of the Kerr-Newman Solution}
\label{Derivation}

In this section, we derive the Kerr-Newman metric for a rotating, charged particle by starting with the Reissner-Nordstr\"om solution for a charged, spherically symmetric particle.  Applying the complex transformation algorithm of \cite{Newman:1965tw, Keane:2014}, we find the metric tensor and Maxwell field of the Kerr-Newman metric, as well as its Weyl tensor components, in a relatively straightforward manner.  The algorithm is then checked explicitly via the spin-coefficient equations.  


\subsection{The complex transformation algorithm}

Our starting point is the Reissner-Nordstr\"om metric for an electrically charged, static, and spherically symmetric body.  In standard spherical coordinates $(ct,r,\theta,\phi)$ the line element for this metric reads:
\be{RN1}
\d s^{2}_{\mathrm{RN}}=\left(1-\frac{r_{\mathrm{S}}}{r}+\frac{r^{2}_{\mathrm{Q}}}{r^2}\right)\;c^2\;\d t^2 -\left(1-\frac{r_{\mathrm{S}}}{r}+\frac{r^{2}_{\mathrm{Q}}}{r^2}\right)^{-1}\;\d r^{2}-r^{2}\;\d\Omega^2.
\ee
Here, the quantities $r_{\mathrm{S}}$ and $r^{2}_{\mathrm{Q}}$ are given by
\begin{equation*}
 r_{\mathrm{S}}=\frac{2GM}{c^2}, \qquad r^{2}_{\mathrm{Q}}=\frac{kG Q^{2}}{c^4},
\end{equation*}
where $c$ is the speed of light, $G$ is Newton's constant, $k$ is Coulomb's constant, and $M$, $Q$ are constant parameters corresponding to the mass and electric charge of the body respectively.  The line element $\d\Omega^{2}$ is the unit sphere metric,
\begin{equation*}
 \d\Omega^{2}=\d\theta^{2}+\sin^{2}\theta \d\phi^{2}.
\end{equation*}
From now on, we simplify notation by working in the natural units $c=G=k=1$ so that $r_{\mathrm{S}}=2M$ and $r^{2}_{\mathrm{Q}}=Q^2$.

We can transform to a null coordinate system where the time coordinate $t$ is replaced with the null time coordinate $u$ via
\begin{equation*}
 u=t-\int_{0}^{r}\frac{r^{\prime\,2}\;\d r^{\prime}}{r^{\prime\,2}-2Mr^{\prime}+Q^2}.
\end{equation*}
In these coordinates, the line element \eqref{RN1} becomes simpler:
\be{RN2}
\d s^{2}_{\mathrm{RN}}=\left(1-\frac{2M}{r}+\frac{Q^2}{r^2}\right)\,\d u^2 +2 \d u\,\d r-r^{2}\;\d\Omega^2.
\ee
Note that the coordinate singularity of \eqref{RN1} in the coefficient of $\d r^2$ is eliminated.  The metric can now be expressed in terms of a \emph{null tetrad} $\{l,n,m,\bar{m}\}$.  These vectors, in turn, become the central variables rather than the metric itself.  The vectors $l,n$ are real while $m$, $\bar{m}$ are complex conjugates; the only non-vanishing scalar products between tetrad vectors are $l\cdot n=1$, $m\cdot\bar{m}=-1$. The contravariant metric is given by
\be{metricrule}
 g^{\mu\nu}=2\left(l^{(\mu}n^{\nu)}-m^{(\mu}\bar{m}^{\nu)}\right).
\ee

From the line element \eqref{RN2}, it is easy to deduce the 1-form duals of the null tetrad:
\begin{eqnarray*}
 l_{\mu}\d x^{\mu} & = & \d u, \\
n_{\mu}\d x^{\mu} & = & \frac{1}{2}\left(1-\frac{2M}{r}+\frac{Q^2}{r^2}\right)\,\d u+\d r, \\
m_{\mu}\d x^{\mu} & = & \frac{r}{\sqrt{2}}\left(\d\theta+i\sin\theta\,\d\phi\right).
\end{eqnarray*}
The inner product and nullity conditions then allow us to write down the tetrad of null vectors itself,
\begin{eqnarray}
 l & = & \frac{\partial}{\partial r}, \label{RNtet1} \\
 n & = & \frac{\partial}{\partial u}-\frac{1}{2}\left(1-\frac{2M}{r}+\frac{Q^2}{r^2}\right)\frac{\partial}{\partial r}, \label{RNtet2} \\
 m & = & \frac{1}{\sqrt{2} r}\frac{\partial}{\partial \theta}+\frac{i\csc\theta}{\sqrt{2}r}\frac{\partial}{\partial \phi} \label{RNtet3} \,.
\end{eqnarray}
The contravariant Reissner-Nordstr\"om metric in null coordinates is thus given by
\be{RN3}
g^{\mu\nu}_{\mathrm{RN}}=\left(
\begin{array}{cccc}
 0 & 1 & 0 & 0 \\
1 & \left(\frac{2M}{r}-\frac{Q^2}{r^2}-1\right) & 0 & 0 \\
0 & 0 & -\frac{1}{r^2} & 0 \\
0 & 0 & 0 & -\frac{\csc^2\theta}{r^2} 
\end{array}\right).
\ee

The electromagnetic potential and Maxwell tensor for the Reissner-Nordstr\"om metric are
\be{RNEM}
A_{\mu}=\left(\frac{Q}{r},0,0,0\right), \qquad F_{\mu\nu}=\left(
\begin{array}{cccc}
 0 & -\frac{Q}{\sqrt{2}r^2} & 0 & 0 \\
 \frac{Q}{\sqrt{2}r^2} & 0 & 0 & 0 \\
0 & 0 & 0 & 0 \\
0 & 0 & 0 & 0
\end{array}\right).
\ee
At this point, we can use \eqref{RNtet1}-\eqref{RNtet3} and \eqref{RNEM} to write down the spin-coefficient expressions for the Maxwell field and Weyl tensor, which are given respectively by
\be{RNMax}
\phi_{0}=F_{\mu\nu}l^{\mu}m^{\nu}=0, \qquad \phi_{1}=\frac{1}{2}F_{\mu\nu}(l^{\mu}n^{\nu}+m^{\mu}\bar{m}^{\nu})=\frac{Q}{\sqrt{2}r^2}, \qquad \phi_{2}=F_{\mu\nu}\bar{m}^{\mu}n^{\nu}=0,
\ee
and
\be{RNWeyl}
\Psi_{0}=-C_{\mu\nu\rho\sigma}l^{\mu}m^{\nu}l^{\rho}m^{\sigma}=0, \qquad \Psi_{1}=-C_{\mu\nu\rho\sigma}l^{\mu}n^{\nu}l^{\rho}m^{\sigma}=0,  
\ee
\begin{equation*}
 \Psi_{2}=-\frac{1}{2}\left(C_{\mu\nu\rho\sigma}l^{\mu}n^{\nu}l^{\rho}n^{\sigma}-C_{\mu\nu\rho\sigma}l^{\mu}n^{\nu}m^{\rho}\bar{m}^{\sigma}\right)=-\frac{M}{r^3}+\frac{Q^2}{r^4},
\end{equation*}
\begin{equation*}
 \Psi_{3}=C_{\mu\nu\rho\sigma}l^{\mu}n^{\nu}n^{\rho}\bar{m}^{\sigma}=0, \qquad \Psi_{4}=C_{\mu\nu\rho\sigma}n^{\mu}\bar{m}^{\nu}n^{\rho}\bar{m}^{\sigma}=0.
\end{equation*}
From the vanishing of $\Psi_{0}$, $\Psi_{1}$, $\Psi_{3}$, and $\Psi_{4}$, the Reissner-Nordstr\"om metric is algebraically special of Petrov type D, with $l$ and $n$ the two DPNVs.  In addition, $l$ and $n$ are also the two principal null directions of the Maxwell field.

\medskip

We can now formalize an algorithm to produce the charged spinning particle metric and fields directly from the Reissner-Nordstr\"om metric and field.  This algorithm is motivated by the similarity between the degenerate null geodesic congruences of the Schwarzschild and Kerr metrics with that of the Reissner-Nordstr\"om metric.  All three metrics are type D, and the principal null geodesic congruence with tangent vector $l$ has complex divergence given by $\rho=-r^{-1}$ for Schwarzschild and Reissner-Nordstr\"om, while for the Kerr metric it is
\begin{equation*}
 \rho=\frac{-1}{r-ia\cos\theta} \,.
\end{equation*}
Since $\rho$ determines almost all the $r$-dependence of solutions in the spin-coefficient formalism, this suggests a transformation $r\rightarrow r+ia\cos\theta$, which was shown to send Schwarzschild to Kerr in \cite{Newman:1965tw}.



Following this observation, we first complexify the null coordinate system by allowing $u$ and $r$ to take values in $\C$.  Inside our expressions for the tetrad vectors \eqref{RNtet1}-\eqref{RNtet3}, this leads to an ambiguity in how to complexify the various functions of $r$.  In general, all that is required is that a function $f(r)$ be promoted to a function $F(r,\bar{r})$ which is rational and reduces to $f$ on the real slice.  One of the simplest such choices is given by rewriting the Reissner-Nordstr\"om tetrad in complexified form
\begin{eqnarray}
l & \rightarrow & l^{\prime}=\frac{\partial}{\partial r}, \label{Ctet1} \\
n & \rightarrow & n^{\prime}=\frac{\partial}{\partial u}-\frac{1}{2}\left(1-\frac{M}{r}-\frac{M}{\bar{r}}+\frac{Q^2}{|r|^2}\right)\frac{\partial}{\partial r}, \label{Ctet2} \\
m & \rightarrow & m^{\prime}= \frac{1}{\sqrt{2} \bar{r}}\left(\frac{\partial}{\partial \theta}+i\csc\theta\frac{\partial}{\partial \phi}\right), \label{Ctet3} \\
\bar{m} & \rightarrow & \bar{m}^{\prime}=\frac{1}{\sqrt{2} r}\left(\frac{\partial}{\partial \theta}-i\csc\theta\frac{\partial}{\partial \phi}\right). \label{Ctet4} 
\end{eqnarray}
  
In this complexified tetrad, we make a complex change of coordinates motivated by the analogy with the Schwarzschild--to--Kerr transformation:
\be{cc1}
r^{\prime}= r+ia\cos\theta, \qquad u^{\prime}= u-ia\cos\theta, \qquad \theta^{\prime}=\theta, \qquad \phi^{\prime}=\phi,
\ee
where $a\in\R$ will later be interpreted as the Kerr parameter.  While the un-primed coordinate system is still viewed as complex, we will interpret the new coordinates as real.  It is a straightforward exercise to translate, by an ordinary coordinate transformation, the complexified tetrad \eqref{Ctet1}-\eqref{Ctet4} into the coordinate system \eqref{cc1}; remarkably, the result is a null tetrad describing a \emph{real} space-time (we drop the primes on the new coordinates to simplify notation):
\begin{eqnarray}
l^{\prime} & = & \frac{\partial}{\partial r}, \label{KNtet1} \\
n^{\prime} & = & \frac{\partial}{\partial u}-\frac{1}{2}\left(1-\frac{2Mr-Q^2}{R^2}\right)\frac{\partial}{\partial r}, \label{KNtet2} \\
m^{\prime} & = & \frac{1}{\sqrt{2} (r+ia\cos\theta)}\left(ia\sin\theta\frac{\partial}{\partial u}-ia\sin\theta\frac{\partial}{\partial r}+\frac{\partial}{\partial \theta}+i\csc\theta\frac{\partial}{\partial \phi}\right), \label{KNtet3} 
\end{eqnarray}
where $R^{2}\equiv r^{2}+a^{2}\cos^{2}\theta$.

This null tetrad allows us to write down the corresponding contravariant metric via \eqref{metricrule}, resulting in
\be{KNmet1}
g^{\mu\nu}=\left(
\begin{array}{cccc}
 -\frac{a^{2}\sin^{2}\theta}{R^2} & \frac{r^2+a^2}{R^2} & 0 & -\frac{a}{R^2} \\
 \cdots & -\frac{\Delta}{R^2} & 0 & \frac{a}{R^2} \\
 \cdots & \cdots & -\frac{1}{R^2} & 0 \\
 \cdots & \cdots & \cdots & -\frac{\csc^{2}\theta}{R^2}
\end{array}\right),
\ee
where we use the definitions
\begin{equation*}
 R^{2}\equiv r^{2}+a^2\cos^{2}\theta, \qquad \Delta\equiv r^{2}+a^2-2Mr+Q^2.
\end{equation*}
It is easy to see that in the $a\rightarrow 0$ limit, $g^{\mu\nu}$ becomes the Reissner-Nordstr\"om metric of \eqref{RN3}.  The covariant form of the metric is given by
\be{KNmet2}
g_{\mu\nu}=\left(
\begin{array}{cccc}
1-\frac{2Mr-Q^2}{R^2} & 1 & 0 & \frac{a\sin^{2}\theta}{R^2}(2Mr-Q^2) \\
\cdots & 0 & 0 & -a\sin^{2}\theta \\
\cdots & \cdots & -R^{2} & 0 \\
\cdots & \cdots & \cdots & \frac{\sin^{2}\theta}{R^2}(\Delta a^{2}\sin^{2}\theta-(a^2+r^2)^2) 
\end{array}\right),
\ee
giving the line element of the Kerr-Newman metric in null coordinates \cite{Newman:1965my}:
\begin{multline}\label{KNmet3}
\d s^{2}_{\mathrm{KN}}=\left(1-\frac{2Mr-Q^2}{R^2}\right)\d u^{2}+2\d u\,\d r+2\frac{a\sin^{2}\theta}{R^2}\left(2Mr-Q^2\right) \d u\,\d\phi \\ -2a\sin^{2}\theta \,\d r\,\d\phi-R^{2}\d\theta^{2}+\frac{\sin^{2}\theta}{R^2}\left(\Delta a^{2}\sin^{2}\theta-(a^2+r^2)^2\right)\,\d\phi^{2}.
\end{multline}

There are some easy consistency checks we can perform on the algorithm so far.  First of all, under $Q\rightarrow 0$ or $a\rightarrow 0$, \eqref{KNmet3} reduces to the Kerr or Reissner-Nordstr\"om metric, respectively.  Furthermore, despite the complexifications used at various steps in the process, the resulting metric is \emph{real}.  Also, the principal null congruence is generated by the vector $l$, whose 1-form dual can be calculated using \eqref{KNmet2} to be $\d u-a\sin^{2}\theta \d\phi$.  This is the analogous congruence to the Kerr metric, just as the Schwarzschild and Reissner-Nordstr\"om metrics share the congruence corresponding to $\d u$.

\medskip

The Kerr-Newman metric is only the gravitational part of a solution to the Einstein-Maxwell equations.  For a full electrovacuum solution, we must also have the associated Maxwell fields.  In the original derivation of the metric, the origins of the appropriate Maxwell field were far from algorithmic: it had to be obtained by directly integrating the Maxwell equations in spin-coefficient form \cite{Janis:1965tx, Newman:1965my}.  Furthermore, the algorithm we have used thus far only operates at the level of the null tetrad, rather than Newman-Penrose quantities themselves (such as the Weyl or Maxwell tensor components).

It seems natural to ask if there is a way in which the algorithm can be modified so that it applies directly to Weyl or Maxwell tensor components.  This would sidestep the issue of integrating the curved-space Maxwell equations, and provide the full Einstein-Maxwell solution algorithmically.  Such a generalization has recently been found by Keane \cite{Keane:2014}, utilizing the Lorentz symmetry associated with the null tetrad \eqref{KNtet1}-\eqref{KNtet3}.

The four tetrad vectors $\{l,n,m,\bar{m}\}$ can undergo the six parameter (local) Lorentz transformations which preserve the metric \eqref{KNmet3} \cite{Newman:1981fn, Newman:2009}.  We make use of a particular member of this set: the null rotation keeping $l$ unchanged while rotating the other vectors about $l$.  It takes the form:
\be{nullrot}
\begin{array}{lll}
 l & \rightarrow l^{*}=l, \\
 n & \rightarrow n^{*}=n+\bar{\alpha}m+\alpha\bar{m}+\bar{\alpha}\alpha l, \\
 m & \rightarrow m^{*}=m+\alpha l,
\end{array}
\ee
where $\alpha$ is a complex function.

Ordinarily the complex pair of null vectors $\{m,\bar{m}\}$ are chosen to be tangent to the surfaces of constant $r$ in space-time.  However, \eqref{KNtet3} shows that the algorithm produces vectors with non-vanishing $r$-component.  Using the null rotation \eqref{nullrot} with 
\be{KNnr}
\alpha=\frac{ia\sin\theta}{\sqrt{2}(r+ia\cos\theta)},
\ee
the $r$-component of $m, \bar{m}$ is eliminated.  The resulting tetrad (dropping the $*$) is:
\begin{eqnarray}
 l & = & \frac{\partial}{\partial r}, \label{KNtet1*} \\
n & = &\left(\frac{r^2+a^2}{R^2}\right) \frac{\partial}{\partial u}-\frac{\Delta}{2R^2}\frac{\partial}{\partial r}+\frac{a}{R^2}\frac{\partial}{\partial\phi}, \label{KNtet2*} \\
m & = & \frac{1}{\sqrt{2} (r+ia\cos\theta)}\left(ia\sin\theta\frac{\partial}{\partial u}+\frac{\partial}{\partial \theta}+i\csc\theta\frac{\partial}{\partial \phi}\right), \label{KNtet3*} 
\end{eqnarray}
which is clearly in the desired form.  

Beginning with the Reissner-Nordstr\"om form of the Weyl and Maxwell tensor components, complexify them as
\begin{equation*}
 \Psi_{0},\Psi_{1},\Psi_{3},\Psi_{4}\rightarrow 0, \qquad \Psi_{2}\rightarrow -\frac{M}{r^3}+\frac{Q^2}{r^{2}|r|^{2}}, \qquad \phi_{0},\phi_{2}\rightarrow 0, \qquad \phi_{1}\rightarrow\frac{Q}{\sqrt{2}r^2} \, .
\end{equation*}
The extended algorithm states that these quantities transform as scalars under the complex coordinate transformation \eqref{cc1}.  In particular, the result for the Maxwell field is 
\be{KNMax}
\phi_{0}=0, \qquad \phi_{1}=\frac{Q}{\sqrt{2}(r-ia\cos\theta)^2}, \qquad \phi_{2}=0,
\ee
and the Weyl tensor components are
\be{KNWeyl}
\Psi_{0}=\Psi_{1}=\Psi_{3}=\Psi_{4}=0, \qquad \Psi_{2}=\frac{-M}{(r-ia\cos\theta)^3}+\frac{Q^2}{(r+ia\cos\theta)(r-ia\cos\theta)^3}.
\ee  
We will justify the validity of this algorithm in the next subsection.


The scalars \eqref{KNMax}, combined with the tetrad \eqref{KNtet1*}-\eqref{KNtet3*}, completely determine the Maxwell field associated with the Kerr-Newman solution.  In particular, the contravariant Maxwell tensor is given by
\be{KNEM}
F^{\mu\nu}=\frac{Q}{R^6}\left(
\begin{array}{cccc}
 0 & (r^4+ a^2 r^2 \sin^{2}\theta -a^{4}\cos^{2}\theta) & -2 a^2 r\cos\theta\sin\theta & 0 \\
 \cdots & 0 & 0 & a(a^2\cos^{2}\theta-r^2) \\
\cdots & \cdots & 0 & 2ar\cot\theta \\
 \cdots & \cdots &  \cdots & 0
\end{array}\right),
\ee
which reduces to the Reissner-Nordstr\"om electric field in the $a\rightarrow 0$ limit.

This completes the complex transformation algorithm, first given in \cite{Newman:1965tw} and extended to the Maxwell and Weyl tensor components in \cite{Keane:2014}.  From \eqref{KNWeyl}, we see that the Kerr-Newman solution is algebraically special of type D; the two repeated principal null directions are in fact given by the tetrad vectors $l$ and $n$ of \eqref{KNtet1*} and \eqref{KNtet2*} respectively.  The algorithm has been the subject of study in its own right (c.f. \cite{Drake:1997hh, Drake:1998gf}), leading to natural coordinate extensions across the singularities of the manifold (see below) \cite{Brauer:2014wwa}, numerical implementations \cite{Gutierrez-Chavez:2014gza}, and it has even found use in alternative theories of gravity (c.f., \cite{Kim:1998iw, CiriloLombardo:2004qw, Hansen:2013owa}).


\subsection{Justification}

This complex transformation algorithm allows us to go from the Reissner-Nordstr\"om metric and associated Maxwell field to the Kerr-Newman metric and Maxwell field without needing to integrate any of the field equations.  However, it is not obvious that the extended algorithm--which acts on the Weyl and Maxwell tensor components directly--should be correct. That is, how do we know that it produces a solution to the Einstein-Maxwell equations, and that the resulting tetrad contains the two real null vectors corresponding to the principal null directions of a type D Weyl and Maxwell tensor?  Here, we justify the extended algorithm by a straightforward (but tedious) calculation via the spin-coefficient (or Newman-Penrose) equations.

Begin with the tetrad \eqref{KNtet1*}-\eqref{KNtet3*}, which is the end-point of the extended algorithm.  Using this tetrad, we calculate all the associated spin-coefficients and verify (by direct substitution) that the Einstein-Maxwell equations are satisfied.  Define the standard differential operators associated with the tetrad:\footnote{We abuse notation by referring to the differential operator $n^{\mu}\partial_{\mu}$ as $\Delta$, along with the algebraic quantity $r^{2}+a^2-2Mr+Q^2$; the distinction should nevertheless be clear from the context.}
\begin{eqnarray*}
 D & \equiv & l^{\mu}\frac{\partial}{\partial x^{\mu}}= \frac{\partial}{\partial r}, \\
 \Delta & \equiv & n^{\mu}\frac{\partial}{\partial x^{\mu}}=\left(\frac{r^2+a^2}{R^2}\right) \frac{\partial}{\partial u}-\frac{\Delta}{2R^2}\frac{\partial}{\partial r}+\frac{a}{R^2}\frac{\partial}{\partial\phi}, \\
\delta & \equiv & m^{\mu}\frac{\partial}{\partial x^{\mu}}=\frac{1}{\sqrt{2} \Gamma}\left(ia\sin\theta\frac{\partial}{\partial u}+\frac{\partial}{\partial \theta}+i\csc\theta\frac{\partial}{\partial \phi}\right),
\end{eqnarray*}
where we abbreviate $\Gamma=r+ia\cos\theta$.  Now, consider the spin-coefficient equations for the metric \cite{Newman:1961qr, Newman:2009}:
\begin{eqnarray}
 \Delta l^{\mu}-Dn^{\mu} & = & (\gamma+\bar{\gamma})l^{\mu}+(\epsilon+\bar{\epsilon})n^{\mu}-(\tau+\bar{\pi})\bar{m}^{\mu}-(\bar{\tau}+\pi)m^{\mu} \\
 \delta l^{\mu}-Dm^{\mu} & = & (\bar{\alpha}+\beta-\bar{\pi})l^{\mu}+\kappa n^{\mu}-\sigma \bar{m}^{\mu}-(\bar{\rho}+\epsilon-\bar{\epsilon})m^{\mu} \\
 \delta n^{\mu}-\Delta m^{\mu} & = & -\nu l^{\mu}+(\tau-\bar{\alpha}-\beta)n^{\mu}+\bar{\lambda}\bar{m}^{\mu}+(\mu-\gamma+\bar{\gamma})m^{\mu} \\
 \bar{\delta}m^{\mu}-\delta\bar{\mu}^{\mu} & = & (\bar{\mu}-\mu)l^{\mu}+(\bar{\rho}-\rho)n^{\mu}-(\bar{\alpha}-\beta)\bar{m}^{\mu}+(\alpha-\bar{\beta})m^{\mu}.
\end{eqnarray}

Following some algebraic manipulations, we obtain expressions for all of the spin-coefficients:
\begin{equation*}
 \kappa=\sigma=\lambda=\nu=\epsilon=0, \qquad \rho=-\frac{1}{\overline{\Gamma}},
\end{equation*}
\begin{equation*}
 \alpha=\frac{\sqrt{2} ia\sin\theta}{2\overline{\Gamma}^2}-\frac{\sqrt{2}\cot\theta}{4\overline{\Gamma}}, \qquad \beta=\frac{\sqrt{2}\cot\theta}{4\Gamma}, \qquad \tau = -\frac{\sqrt{2} ia\sin\theta}{2\,\Gamma\,\overline{\Gamma}},
\end{equation*}
\begin{equation*}
 \mu=-\frac{\Delta}{2\,\Gamma\,\overline{\Gamma}^2}, \qquad \gamma=-\frac{\Delta}{2\,\Gamma\,\overline{\Gamma}^2}+\frac{r-M}{2\,\Gamma\,\overline{\Gamma}}, \qquad \pi=\frac{\sqrt{2}ia\sin\theta}{2\overline{\Gamma}^2}.
\end{equation*}
At this point, the vanishing of the spin-coefficients $\kappa$, $\sigma$, $\lambda$, and $\nu$ implies that the Weyl and Maxwell tensor components
\begin{equation*}
 \Psi_{0}=\Psi_{1}=\Psi_{3}=\Psi_{4}=\phi_{0}=\phi_{2}=0,
\end{equation*}
by the Goldberg-Sachs theorem \cite{Goldberg:1962} (which is expressed particularly elegantly in the spin-coefficient formalism \cite{Newman:1961qr}).

The remaining Weyl and Maxwell tensor components are determined by `field equations' for the spin-coefficients themselves.  In particular, $\Psi_{2}$ is determined by
\begin{equation*}
 D\mu-\delta\pi=\bar{\rho}\mu+\pi\bar{\pi}-\pi(\bar{\alpha}-\beta)+\Psi_{2},
\end{equation*}
which we can solve for $\Psi_{2}$ by substituting the expressions for the spin-coefficients above,
\be{Weyl*}
\Psi_{2}=-\frac{M}{\bar{\Gamma}^3}+\frac{Q^2}{\Gamma\bar{\Gamma}^3}.
\ee
Likewise, the Maxwell tensor component $\phi_{1}$ is determined by the field equation for $\gamma$,
\begin{equation*}
 D\gamma=(\tau+\bar{\pi})\alpha+(\bar{\tau}+\pi)\beta+\tau\pi+\Psi_{2}+\phi_{1}\bar{\phi}_{1},
\end{equation*}
leading to
\be{Max*}
\phi_{1}=\frac{Q}{\sqrt{2}\bar{\Gamma}^2}.
\ee

Clearly, the expressions \eqref{Weyl*} and \eqref{Max*} (which are derived directly from the field equations) agree with \eqref{KNWeyl} and \eqref{KNMax} produced by the extended complex transformation algorithm.  This confirms that the second portion of the algorithm, first presented in \cite{Keane:2014}, is correct and the Maxwell and Weyl tensor components transform as scalars.


\subsection{Extensions}

At this point, it seems natural to ask if the `complexification trick' underlying this algorithm can be extended to other space-times.  It turns out that the Kerr-Newman metric is an example of a broad class of metrics known as \emph{Kerr-Schild} metrics \cite{Kerr:1965}.  Besides being of physical interest (for instance, all four black hole solutions are Kerr-Schild), complexification acts naturally on a very large sub-class of these metrics, as we explain below.

Generally, a Kerr-Schild metric is defined by a scalar function $\mathcal{F}$ and 1-form $l_{\mu}\d x^{\mu}$ via
\be{KS1}
g_{\mu\nu}=\eta_{\mu\nu}+\mathcal{F}\;l_{\mu}l_{\nu},
\ee
where $\eta_{\mu\nu}$ is the Minkowski metric and $l_{\mu}$ is required to be null with respect to $g$ (and hence also with respect to $\eta$).  The nullity of $l$ greatly simplifies the field equations for this class.  Furthermore, when $l$ is a geodesic field the situation becomes even better: the space-time must be algebraically special of Petrov type II or D.  The Einstein-Maxwell equations for a general metric of Kerr-Schild form were first studied in \cite{Debney:1969zz}, where explicit solutions (including Kerr and Kerr-Newman) were obtained as special cases.  The utility of the Kerr-Schild ansatz is particularly apparent for the field equations in the spin-coefficient formalism: provided the stress tensor obeys some simple restrictions, all but five of the spin-coefficients can be set to zero.  Those that remain are determined by four independent complex functions, which are themselves constrained by a dramatically simplified set of field equations \cite{Debever:1974, Gurses:1975vu}.

There is a sub-class of Kerr-Schild metrics for which the field equations are simplified even further.  While the precise definition of this sub-class is a bit technical (relying on the behavior of functions determining the twist of the null vector $l$), the reduced field equations are remarkably compact.  In a null coordinate system $(u,r,\zeta,\bar{\zeta})$, with $(\zeta,\bar\zeta)$ standard stereographic coordinates on the sphere, the field equations are \cite{Talbot:1969}
\begin{eqnarray}
 \Psi^{0}_{2} & = & c\,u +a(\zeta,\bar\zeta)+i b(\zeta,\bar\zeta), \label{talbot1}\\
 2i b(\zeta,\bar\zeta) & = & P^{2}\frac{\partial^{2}}{\partial\zeta\partial\bar\zeta}\left[P^2\left(\frac{\partial \bar{Q}}{\partial\bar\zeta}-\frac{\partial Q}{\partial\zeta}\right)\right]+2P^{4}\left(\frac{\partial \bar{Q}}{\partial\bar\zeta}-\frac{\partial Q}{\partial\zeta}\right)\frac{\partial^{2}}{\partial\zeta\partial\bar\zeta}\ln P, \\
 0 & = & \frac{\partial}{\partial\bar\zeta}\left(a(\zeta,\bar\zeta)+i b(\zeta,\bar\zeta)\right) \label{talbot3}.
\end{eqnarray}
Here, $\Psi^{0}_{2}$ is the leading coefficient in the peeling expansion of $\Psi_{2}$ and a function of $(u,\zeta,\bar\zeta)$ alone; $c$ is a real constant; $a$, $b$, and $P$ are real-valued functions on the sphere, and $Q$ is a complex-valued function on the sphere.  Remarkably, all known examples of Kerr-Schild solutions to the field equations lie in the sub-class defined by these field equations \cite{Robinson:1969}.

Talbot has shown that a generalized version of the complexification algorithm transforms one set of solutions to \eqref{talbot1}-\eqref{talbot3} to another solution \cite{Talbot:1969}.  In particular, suppose that we have a given solution in terms of various functions of $(u,r,\zeta,\bar\zeta)$, and we perform the complex transformation
\begin{equation*}
 r^{\prime}=r+iT(\zeta,\bar\zeta), \qquad u^{\prime}=u+iS(\zeta,\bar\zeta),
\end{equation*}
where $S$ and $T$ are real-valued.  Then the result will also define a solution of the field equations (in the primed coordinates, interpreted as real-valued), with
\begin{equation*}
 \Psi^{0\;\prime}_{2}=c\,u^{\prime}+a+a_{0}+i(b-b_{0})+i S\,c,
\end{equation*}
for $a_{0},\,b_{0}$ real constants, provided the functions $S$ and $T$ satisfy:
\begin{equation*}
 b_{0}-S\,c+P^{2}\frac{\partial^{2} T}{\partial\zeta\partial\bar\zeta}+2TP^2\frac{\partial^{2} \ln P}{\partial\zeta\partial\bar\zeta}=0, \qquad P^2 \frac{\partial^{2} S}{\partial\zeta\partial\bar\zeta}=T.
\end{equation*}
It is easy to see that \eqref{cc1} is an example of such general complex transformations.  Hence, the basic `trick' of \cite{Newman:1965tw} can be explained for a (large and physically interesting) sub-class of Kerr-Schild solutions at the level of the field equations: any solution to the reduced field equations is transformed to some other solution by an admissable complex coordinate transformation.  

The sort of solutions which can be obtained in this way include some Robinson-Trautman, electrovacuum, and perfect fluid metrics (c.f., \S 29.2 of \cite{Stephani:2003}).  However, it is important to note that Talbot's result holds in principal as well as in practice: we don't actually need the explicit solution in order to know that the complex transformation will produce another solution.

Demia\'nski found a class of Petrov type II Einstein-Maxwell metrics using a complex coordinate transformation by starting from a generalization of the complexified Reissner-Nordstr\"om solution \cite{Demianski:1972uza}.  A cosmological constant can be included for these metrics \cite{Quevedo:1992}, making them the most general known solutions which can be obtained by means of the complexification algorithm, although they lack an attractive physical interpretation due to `wire singularities' extending to infinity.  This demonstrates the limitations of such complex transformation algorithms: the most general type D vacuum (plus cosmological constant) solution to the Einstein-Maxwell equations \cite{Plebanski:1976gy} cannot be obtained in this way.  

Finally, we note that the complexification technique can be understood in another, slightly orthogonal, way for several interesting Kerr-Schild solutions to the field equations.  Under certain assumptions on the null 1-form $l_{\mu}\d x^{\mu}$, a Kerr-Schild solution can be generated from a single potential function $\Upsilon$ on flat $\R^3$, which is harmonic and whose inverse solves the Eikonal equation \cite{Schiffer:1973}:
\begin{equation*}
 \nabla^{2}_{\R^3}\Upsilon=0, \qquad (\nabla_{\R^3}\Upsilon)^2=\Upsilon^{4}.
\end{equation*}
The Schwarzschild metric is an example of such a solution, generated by $\Upsilon=(x^2+y^2+z^2)^{-1/2}$.  Performing the complex transformation $z\rightarrow z-ia$ generates a different solution: the Kerr metric \cite{Schiffer:1973}!  This approach of complexifying generating potentials has been extended to the charged setting \cite{Finkelstein:1974nr, Collins:1976}, incorporating the Reissner-Norstr\"om and Kerr-Newman metrics.  Although its precise relationship to the complex transformation algorithm discussed above remains unclear, an explanation does exist in Minkowski space \cite{Newman:2002mk}.


\section{Properties of the Solution}
\label{Properties}

In this section, we discuss some interesting geometric features of the Kerr-Newman metric.  First, however, we display the metric (given by \eqref{KNmet3} in a null coordinate system) in two other frequently used coordinate systems: Boyer-Lindquist and Kerr-Schild.  This is followed by a discussion of its singularities, horizons, and ergosphere.  A fairly detailed treatment of these topics in the context of the Kerr metric can be found in \cite{Visser:2007fj}. 

First, we remark that the interpretation of the three parameters $M, Q, a$ appearing in the solution is carried over from their meaning in the Schwarzschild, Reissner-Nordstr\"om, and Kerr solutions respectively.  The parameter $M$ is identified as the mass, $Q$ as the electric charge, and $a$ as the angular momentum per unit mass (or Kerr parameter).  Note that these interpretations can also be obtained using the asymptotic definitions of mass, angular momentum, and charge (c.f., \cite{Newman:1981fn, Szabados:2009eka, Adamo:2009fq, Adamo:2009vu}).

\subsection{Alternate coordinate systems}

\subsubsection*{\textit{Boyer-Lindquist coordinates}}

In the null coordinate system in which we have worked so far, there is a proliferation of off-diagonal entries in the metric.  Besides making life difficult from a calculational point of view, it would be useful to have an explicit Lorentzian time coordinate when trying to understand the event horizons and ergospheres of the solution later on.  There is a coordinate system which minimizes the off-diagonal components and introduces an explicit time-like coordinate, namely \emph{Boyer-Lindquist} coordinates, which were originally introduced for the Kerr metric \cite{Boyer:1966qh}.

The transformation to Boyer-Lindquist coordinates is given by the 1-forms \cite{Carter:1968rr}
\be{BL1}
\d t=\d u+\frac{r^{2}+a^{2}}{\Delta}\d r, \qquad \d\varphi = \d\phi +\frac{a}{\Delta}\d r.
\ee
A lengthy but straightforward algebraic manipulation then leads to the new form of the line element:
\be{BL2}
\d s^{2}=\frac{\Delta}{R^2}\left(a\sin^{2}\theta \d\varphi-\d t\right)^{2}-\frac{\sin^{2}\theta}{R^2}\left((r^2+a^2)\d\varphi -a\d t\right)^{2} -\frac{R^2}{\Delta}\d r^2 -R^{2}\d\theta^{2}.
\ee
The only off-diagonal component of the metric in this new coordinate system is $g_{t\varphi}$.  Furthermore, $t$ is explicitly a time coordinate, as can be seen upon expanding the metric for large $r$:
\begin{multline*}
 \d s^{2}=\left(1-\frac{2Mr-Q^2}{r^2}+O(r^{-3})\right)\d t^{2}+\left(4\frac{Ma\sin^{2}\theta}{r}+O(r^{-2})\right)\d t\, \d\varphi \\
-\left(1-\frac{2Mr-Q^{2}}{r^2}+O(r^{-3})\right)\left(\d r^{2}+r^{2}\d\Omega^{2}\right).
\end{multline*}
Note that this asymptotic expansion demonstrates another advantage of the Boyer-Lindquist coordinates: the metric is manifestly asymptotically flat, and our physical identifications for the parameters $M, Q,$ and $a$ are confirmed.

There is another interesting practical consequence of working in Boyer-Lindquist coordinates: the Hamilton-Jacobi equation governing the motion of a test particle is now separable.  This leads to the discovery of a fourth constant of motion (in addition to energy, angular momentum, and rest mass) associated with the Kerr-Newman metric, and completely characterizes the geodesic motion of test particles in the space-time.  For a test particle of mass $\mu$, energy $E$, three-momentum $\vec{p}$, and axial angular momentum $L$, this quantity--known as \emph{Carter's constant}--is given by \cite{Carter:1968rr}:
\be{Carter}
\mathcal{C}=p_{\theta}^{2}+\cos^{2}\theta\left[a^{2}\left(\mu^2-E^2\right)+\left(\frac{L}{\sin\theta}\right)^2\right],
\ee  
where $p_{\theta}$ is the component of $\vec{p}$ in the $\theta$-direction.

\subsubsection*{\textit{Kerr-Schild coordinates}}

We have already discussed the fact that the Kerr-Newman metric is an example of a larger class known as Kerr-Schild metrics, having the form \eqref{KS1}.  Expressing the metric in this way has the advantage of allowing us to explicitly see the 1-form dual to the principal shear-free null geodesic congruence associated with the space-time by the Goldberg-Sachs theorem.  Furthermore, the role of mass and charge in `shifting' the geometry away from flat Minkowski space becomes explicit.

It turns out that our original null coordinate description \eqref{KNmet3} is already in Kerr-Schild form, albeit in a rather non-obvious coordinate system.  To see this, we can pull all of the $M,Q$ dependence of the metric into a single term of the line element so that \eqref{KNmet3} becomes:
\begin{multline}\label{KSform1} 
\d s^{2}=\left(\d u+a\sin^{2}\theta \d\phi\right)^{2}+2\left(\d u-a\sin^{2}\theta \d\phi\right)\left(\d r-a\sin^{2}\theta\d\phi\right)-R^{2}\d\Omega^{2} \\
-\left(\frac{2Mr-Q^2}{R^2}\right)\left(\d u-a\sin^{2}\theta \d\phi\right)^{2}.
\end{multline}
In other words, the metric is decomposed as
\begin{equation*}
 g_{\mu\nu}=g^{*}_{\mu\nu}+\mathcal{F}\;l_{\mu}\,l_{\nu},
\end{equation*}
with
\be{Mink1}
g^{*}_{\mu\nu}\d x^{\mu} \d x^{\nu}=\d u^{2}+2 \d u\,\d r-R^{2}\d\theta^{2}-\sin^{2}\theta\,(a^2+r^2)\d\phi^{2}-2a\sin^{2}\theta\,\d r\,\d\phi,
\ee
\be{KSform2}
\mathcal{F}\;l_{\mu}\,l_{\nu}\d x^{\mu}\,\d x^{\nu}=-\left(\frac{2Mr-Q^2}{R^2}\right)\left(\d u-a\sin^{2}\theta \d\phi\right)^{2}.
\ee

While it certainly does not appear so at first glance, $g^{*}_{\mu\nu}$ is actually the metric of Minkowski space in an unusual choice of coordinates, known as \emph{oblate spheroidal coordinates}.  The 1-form
\begin{equation*}
 l_{\mu}\d x^{\mu}=\d u-a\sin^{2}\theta \d\phi,
\end{equation*}
is null with respect to both the full Kerr-Newman metric and the flat metric $g^{*}_{\mu\nu}$.  We thus have the result that \eqref{KNmet3} is in Kerr-Schild form, but not with respect to standard Minkowski space coordinates.

The transformation from the traditional flat space null polar coordinate system
\begin{equation*}
 \d s^{2}=\d u^{\prime\,2}-2\d u^{\prime}\,\d r^{\prime}-r^{\prime\,2}\d\Omega^{\prime 2}
\end{equation*}
to that of \eqref{Mink1} can be obtained via the coordinate transformation \cite{Newman:2002mk}:
\begin{eqnarray}
 r^{\prime\,2} & = & r^2+a^2\sin^{2}\theta, \label{oblate1} \\
 \cos^{2}\theta^{\prime} & = & \frac{r^2 \cos^{2}\theta}{r^2+a^2\sin^{2}\theta}, \\
 u^{\prime} & = & u+r -\sqrt{r^2+a^2\sin^{2}\theta}, \\
\phi^{\prime} & = & \phi-\arctan \frac{r}{a} \label{oblate4}.
\end{eqnarray}
This can be further continued to the cartesian Minkowski coordinates by
\begin{eqnarray}
 x & = & r^{\prime}\sin\theta^{\prime}\cos\phi^{\prime} = (r\sin\phi+a\cos\phi)\sin\theta, \\
 y & = & r^{\prime}\sin\theta^{\prime}\sin\phi^{\prime} = (a\sin\phi-r\cos\phi)\sin\theta, \\
 z & = & r^{\prime}\cos\theta^{\prime} = r\cos\theta, \\
 t & = & u+r.
\end{eqnarray}

Performing these transformations to \eqref{KNmet3}, we recover the conventional Kerr-Schild form of the Kerr-Newman metric \cite{Debney:1969zz}:
\be{KS2}
 \d s^{2}=\d t^{2}-\d x^{2}-\d y^{2}-\d z^{2} +\mathcal{F} \left(l_{\mu}\d x^{\mu}\right)^{2},
\ee
where
\begin{equation*}
 \mathcal{F}=-\frac{2Mr^{3}-Q^{2}r^{2}}{r^4+a^2 z^2}
\end{equation*}
and
\begin{equation*}
 l_{\mu}\d x^{\mu}=\d t+\frac{z}{r}\d z+\frac{r}{r^2+a^2}(x \d x+y\d y)-\frac{a}{r^2+a^2}(x\d y-y\d x),
\end{equation*}
for $r$ defined implicitly with respect to $(x,y,z)$ by the algebraic relation
\begin{equation*}
 \frac{x^2+y^2}{r^2+a^2}+\frac{z^2}{r^2}=1.
\end{equation*}
In these coordinates, the complex divergence of $l^{\mu}=l_{\nu}g^{\mu\nu}$ becomes
\begin{equation*}
 \rho=\frac{-1}{r-ia\cos\theta}=\frac{-r}{r^{2}-iar^{\prime}\cos\theta^{\prime}},
\end{equation*}
while the vector potential associated with the Kerr-Newman Maxwell field is
\be{KSMax}
A_{\mu}=\frac{Q r^3}{r^4+a^{2}z^{2}}\left(1,\frac{rx+ay}{r^2+a^2},\frac{ry-ax}{r^2+a^2}, \frac{z}{r}\right).
\ee


\subsection{Source and singularity}

A simple inspection of the Kerr-Newman metric (in any coordinate system) reveals the presence of singularities, which are essentially the same as those of the Kerr metric.  For instance, the metric component $g_{uu}$ in \eqref{KNmet3} becomes singular whenever $R^{2}=0$; equivalently, the expressions for the Weyl and Maxwell tensors in \eqref{KNWeyl}, \eqref{KNMax} shows that both fields are singular at the values
\be{sing1}
r=0, \qquad \theta =\frac{\pi}{2}.
\ee
This can be seen to be an honest geometric singularity by computing curvature contractions (e.g., $R_{\mu\nu\rho\sigma}R^{\mu\nu\rho\sigma}$).  Further, it is an extended, rather than `point-like' (as in Schwarzschild) singularity, as can be seen by the oblate spheroidal coordinate transformation \eqref{oblate1}-\eqref{oblate4} to Kerr-Schild form discussed above.

In particular, $r=0$, $\theta=\pi/2$ in the original coordinates gives
\begin{equation*}
 r^{\prime}=a, \qquad \theta^{\prime}=\frac{\pi}{2},
\end{equation*}
in the associated Minkowski space, so the singularity is on a circle of radius $a$ around the origin in the $z=0$ plane.  Hence, the source of the solution can be considered to lie uniformly distributed on this circle, bounding an interior disc $r^{\prime}<a$ corresponding to $r=0$.  The presence of this singularity supports the interpretation of the Kerr-Newman metric as describing a \emph{black hole} with angular momentum and charge.

This disc immediately raises a problem, however.  Consider approaching the disc along the symmetry axis ($\theta=0$) from \emph{positive} values of $r$, passing through the disc ($r=0$), and then continuing on to \emph{negative} values of $r$ without intersecting the singular ring.  But \eqref{KNWeyl}, \eqref{KNMax} indicate that for large negative values of $r$ (i.e., $r<<0$) the effective values of $M$, $Q$, and $a$ will all change sign.\footnote{More precisely, the Bondi mass aspect, which encodes the asymptotic definition of mass, will change sign upon passing through the disc \cite{Israel:1976vc, Israel:1990hm}.}  This indicates that we are dealing with a double-sheeted manifold with branch cut on the disc bounded by the singular ring; the two branches correspond to $r>0$ and $r<0$.

To avoid this double-sheeted manifold (or double-valuedness in the observable mass, charge, and angular momentum), an additional distributional source for both the matter and charges must be placed on the branch cut.  This results in a membrane, so that approaching the disc from both sides gives a discontinuity in the normal derivatives across it.  The details of these distributional sources are rather complicated, so we will not present them here; a primary reference for both the matter and charged sources is \cite{Israel:1970kp}.  In the special case of vanishing gravitational constant, we are dealing with a Maxwell field in Minkowski space whose source is a rotating charged disc with boundary.  This electromagnetic field has many fascinating properties, and has been studied in its own right (c.f., \cite{LyndenBell:2004fk, Kaiser:2001yn}).


\subsection{Horizons, ergosurfaces, and the ergosphere}

Besides the ring-like curvature singularity, there is additional `singular behavior' for the metric components in the various coordinate systems we have discussed.  In actuality, these additional singularities can be removed by coordinate transformations, so they do not represent actual physical curvature singularities in space-time.  Nevertheless, such \emph{coordinate singularities} often underlie important structures which are of physical interest and have geometric descriptions independent of the choice of coordinates.

Consider the Boyer-Lindquist expression for the Kerr-Newman metric in \eqref{BL2}.  In these coordinates, there appears to be an additional singularity besides the curvature singularity at $R^{2}=0$; namely, the metric component
\begin{equation*}
 g_{rr}=-\frac{R^{2}}{\Delta},
\end{equation*}
becomes singular when $\Delta=0$.  This determines two three-surfaces of constant $r$ given by the roots of $\Delta=0$, namely
\be{EH1}
r=r_{\pm}=M\pm\sqrt{M^{2}-a^2-Q^2}.
\ee
These surfaces are referred to as the outer ($r_{+}$) and inner ($r_{-}$) horizons; the former is called the \emph{event horizon}, and the region $r<r_+$ is referred to as the `interior' of the black hole.  Note that it is easy to see that the singularities in the Boyer-Lindquist form of the metric associated with $r_{\pm}$ are coordinate singularities: indeed, the null coordinate expression for the metric \eqref{KNmet3} has no singularity when $\Delta=0$. 

Despite the fact that the singularity associated with the event horizon is a coordinate artefact, it nevertheless has an important geometric meaning which can be seen via the following argument.  Though not immediately obvious, the interior of the event horizon cannot be accessed in the coordinate system of \eqref{KNmet3}, which is based on the `retarded' null surfaces $u=\mathrm{const.}$ (c.f., \cite{Carter:1968rr, Hawking:1973}).  Instead, we must make use of `advanced' null surfaces based on a coordinate $v=\mathrm{const.}$.  Since the Kerr-Newman metric is stationary and axisymmetric, this can be done immediately by replacing $u$ with $-v$ and $\phi$ with $-\phi$ in \eqref{KNmet3}, leading to
\begin{multline}\label{KNadv}
 \d s^{2}=\left(1-\frac{2Mr-Q^2}{R^2}\right)\d v^{2}-2\d v\,\d r+2\frac{a\sin^{2}\theta}{R^2}\left(2Mr-Q^2\right) \d v\,\d\phi \\ +2a\sin^{2}\theta \,\d r\,\d\phi-R^{2}\d\theta^{2}+\frac{\sin^{2}\theta}{R^2}\left(\Delta a^{2}\sin^{2}\theta-(a^2+r^2)^2\right)\,\d\phi^{2},
\end{multline}
with $R^2$ and $\Delta$ being given by the same expressions as before.

Making the same replacements in the tetrad vectors $l, n$ would lead to past-pointing null vectors; to avoid this we simply change signs and then apply the substitution.  So in the advanced null coordinates, \eqref{KNtet1*} and \eqref{KNtet2*} become
\begin{eqnarray}
 l & = & -\frac{\partial}{\partial r}\, , \label{KNav1} \\
n & = & \left(\frac{r^2+a^2}{R^2}\right) \frac{\partial}{\partial v}+\frac{\Delta}{2R^2}\frac{\partial}{\partial r}+\frac{a}{R^2}\frac{\partial}{\partial\phi}\,.\label{KNav2}
\end{eqnarray}
Note that along a null geodesic approaching the source from \emph{past null infinity}, the vector $l$ is inward and future-pointing (i.e., $\partial_r$ is outwards and past-pointing), and $n$ is outward and future-pointing until the event horizion, defined by the 3-surface $r=r_+$ where $\Delta=0$.  Inside the event horizon, the change of sign in \eqref{KNav2} results in $n$ pointing inward from the surface.

Following the null geodesic as $r$ approaches $r_+$, the future-pointing null cones with apex on the geodesic tip towards the curvature singularity until $r=r_+$.  At that point, one null vector on the cone ($n$) becomes tangent to the event horizon while all other null rays on the cone point \emph{inside} the horizon.  This is true since each ray on the cone can be written as a linear combination of $l$ and $n$ from \eqref{KNav1}--\eqref{KNav2}.

Hence, within the event horizon no physical (null or time-like) trajectories are capable of returning to the exterior ($r>r_+$) of the space-time.  This is manifested by the coordinate $r$ becoming time-like inside the event horizon.  This means that no observer has access to the interior of the black hole, so we will not discuss the inner horizon $r_-$.  We avoid any discussion of possible quantum effects.    

The Kerr-Newman metric is referred to as `extremal' if $M^{2}-a^{2}-Q^{2}=0$; in this case the inner and outer horizon coincide at $r=M$.  If $M^{2}-a^{2}-Q^{2}<0$, then there is no real solution to $\Delta=0$, so there is no event horizon to hide the curvature singularity from exterior observers in the space-time.  Such a situation seems to lead to a naked singularity, in violation of the cosmic censorship conjecture \cite{Penrose:1969pc}, and we will not discuss this super-extremal case here, although there is a wealth of interesting work on the subject.

\medskip

Another surface of physical and geometric importance arises when considering the $g_{tt}$ component of the metric in Boyer-Lindquist coordinates.  Introduce a static world-line, which has constant $(r,\theta,\varphi)$, so that the metric restricted to the world-line is
\be{gtt}
 \d s^{2}|_{r,\theta,\varphi=\mathrm{const.}}=g_{tt}\,\d t^{2}=R^{-2}\left(r^2+a^2\cos^2\theta-2Mr+Q^{2}\right)\;\d t^{2}.
\ee
The outermost three-surface determined by $g_{tt}=0$ is then given explicitly by
\be{ergo}
r=r^{E}_{+}(\theta)=M+ \sqrt{M^2-a^2\cos^2\theta-Q^2},
\ee
where we ignore the other solution $r^{E}_{-}(\theta)$ since it lies inside the event horizon.  From \eqref{gtt}, we see that for $r>r^{E}_{+}(\theta)$, the tangent to the static world-line $k=\partial_{t}$ is time-like, while for $r<r^{E}_{+}(\theta)$ the world-line is space-like.  The surface \eqref{ergo} is often referred to as the stationary limit surface or \emph{ergosurface}, and the region between the ergosurface and event horizon $r_{+}<r<r^{E}_{+}(\theta)$ as the \emph{ergosphere}.  

The invariant geometric meaning of the ergosphere is the region of space-time where the time-translation Killing vector $k=\partial_t$ becomes space-like.  An additional geometric effect is that local light cones are tilted within the ergosphere so that every time-like vector acquires a rotational component (i.e., a $\varphi$-component).  Hence, no physical trajectory can remain stationary within the ergosphere in any time-independent coordinate system.

\begin{figure}[t]
\centering
\includegraphics[width=3.3 in, height=2.5 in]{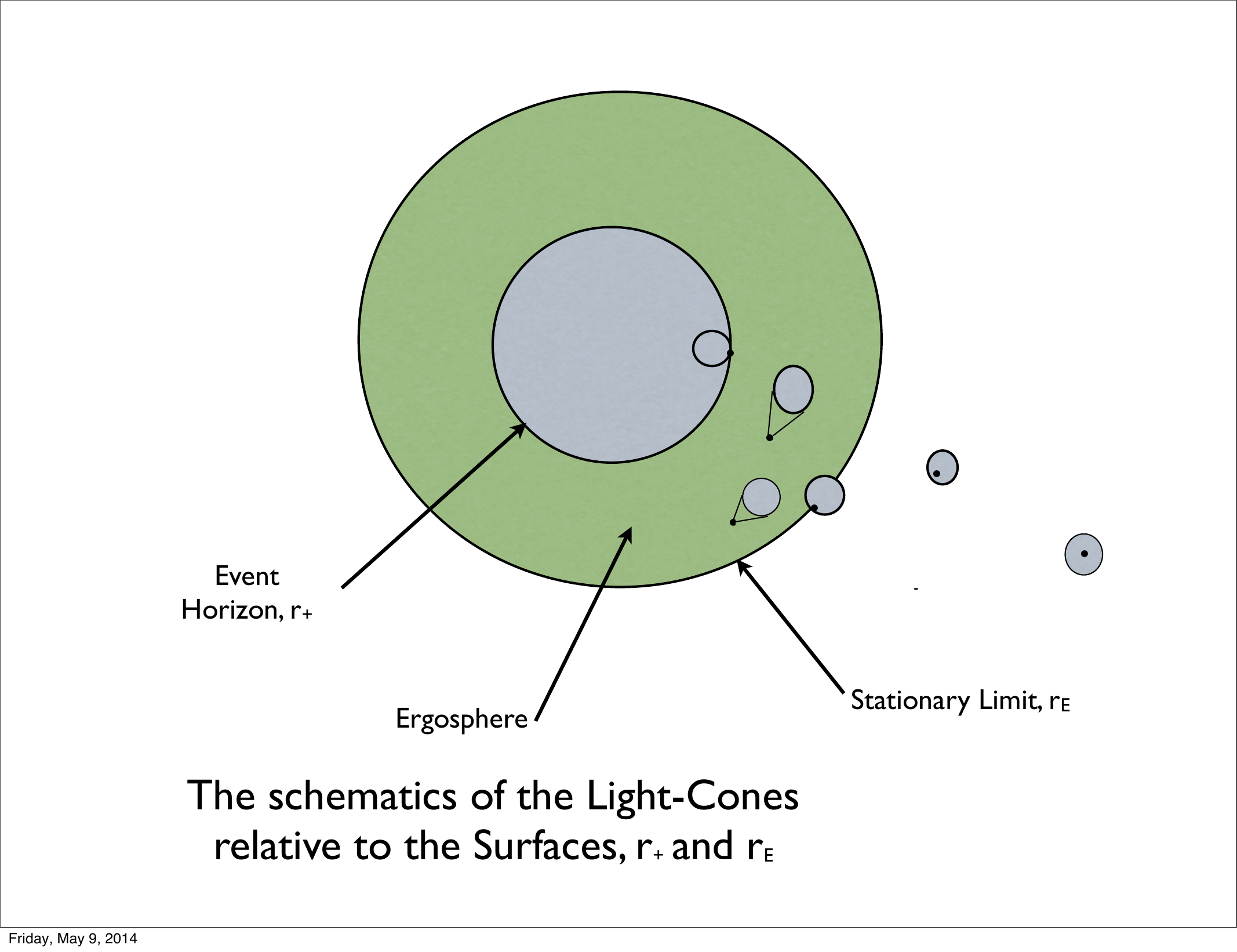}\caption{\small{\textit{Light cones in and around the ergosphere and event horizon}}}\label{Surfaces}
\end{figure}

A useful visualization of many of the properties associated with the event horizon and ergosurface is to consider the local light cones in the different regions of space-time, as illustrated in figure \ref{Surfaces}.  Outside the ergosurface, the light cones are `normal', while at the ergosurface they are tilted in the direction of black hole rotation with two null rays tangent to the surface.  In the ergosphere, the light cone is further tilted so that all time-like rays have components in the direction of rotation.  Finally, at the event horizon the gravitational pull of the black hole has tilted the light cone so that only one null ray lies tangent to the $r=r_{+}$ surface; all other null and time-like directions point into the black hole interior.  
  

\section{Further Advances}
\label{Advances}

We conclude this review with a short discussion of further concepts and advances related to the Kerr-Newman metric.  This is hardly an exhaustive exposition, and the reader will certainly find other interesting and important topics elsewhere in the literature.  Here we focus on: interpretive aspects of the Kerr-Newman metric; additional related solutions of general relativity; and analogues in higher dimensions.

\subsubsection*{\textit{Interpretive issues}}

It was first noted by Carter \cite{Carter:1968rr} that the gyromagnetic ratio associated with the Kerr-Newman metric is that of the Dirac electron.  The classical gyromagnetic ration $\gamma_{\mathrm{class}}$, which is defined by the ratio of the magnetic moment to the (spin) angular momentum, turns out to be $\gamma_{\mathrm{class}}=\frac{Q}{2 M}$ for simple systems.  The Dirac theory of the electron yields a gyromagnetic ratio that is twice as large: $\gamma_{\mathrm{Dirac}}=g\gamma_{\mathrm{class}}$, with $g=2$.  From the asymptotic magnetic field of the Kerr-Newman solution, one sees that its magnetic dipole moment is given by $\mu=Qa$, while the angular momentum is $J=Ma$.  Hence, we immediately recover $\gamma_{\mathrm{KN}}=Q/M$, so $g=2$ and the Dirac value is obtained.

This has led several authors to speculate that the Kerr-Newman black hole could provide some sort of classical `model' for the Dirac electron (c.f.,\cite{Israel:1970kp, Burinskii:2005mm}).  It remains unclear the extent to which this proposal can be fully realized.  One potential issue is the fact that the Kerr-Newman metric also has an associated electric quadrupole moment, which cannot be incorporated into the Dirac theory and has no basis in experimental observation. However, in more general asymptotically flat space-times, $g=2$ can still be recovered when the centers of mass and charge coincide \cite{Kozameh:2008gw, Adamo:2009vu, Adamo:2011cx}.

\subsubsection*{\textit{Energy and thermodynamics}}

Studying the energy associated with the various black hole solutions also leads to interesting consequences.  There is no unique definition of energy for a system in curved space-time, but various workable definitions exist.  For the Schwarzschild solution, the definitions due to Landau and Lifshitz \cite{Landau:1951}, Tolman \cite{Tolman:1930zz}, and Penrose \cite{Penrose:1982wp} result in the energy being concentrated inside the event horizon \cite{Vaidya:1952zz}, while for Reissner-Nordstr\"om the energy is distributed in both the exterior and interior \cite{Tod:1983zx}.

For the Kerr-Newman black hole, Virbhadra calculated the energy contained in the region $r>R$ for some finite $R>r_{+}$ using both the Tolman and Landau-Lifshitz definitions \cite{Virbhadra:1990vs}.  The result is the same in both cases, and is given by:
\begin{equation*}
 E=M-\frac{Q^2}{R}\left(\frac{a^2}{3R^2}+\frac{1}{2}\right).
\end{equation*}
The mass term is concentrated inside the horizon, just as the Schwarzschild case.  This allows us to infer the following fact about the energy distribution of the Kerr space-time: when the charge $Q$ is set to zero, all of the energy lies in the black hole interior.  Likewise, setting $a=0$ yields the energy formulae of \cite{Vaidya:1952zz, Tod:1983zx} for the Reissner-Nordstr\"om metric.  Hence, we see that the energy (suitably defined) of any uncharged black hole is concentrated within the event horizon, while charged black holes have non-vanishing energy density in the exterior as well.  

In this context Penrose \cite{Penrose:1969pc} suggested that it is possible to extract energy from a Kerr or Kerr-Newman black hole by dropping a piece of matter into the ergosphere and letting it split into two parts.  Within the ergosphere, space itself is rotating, so if one portion of matter falls through the event horizon and into the black hole while the other exits the ergosphere and escapes to infinity, the latter portion will extract rotational energy with it.  In this way, it is possible for a rotating black hole to eventually lose all of its angular momentum; the precise amount of energy that can be lost by any black hole due to the Penrose process was calculated in \cite{Christodoulou:1970wf}. 

The amount of energy remaining in a black hole after using this--or any other energy extraction--process is captured by a quantity called the \emph{irreducible mass} of a black hole.  This was first analyzed for a Kerr-Newman black hole by Christodoulou and Ruffini \cite{Christodoulou:1972kt}, who showed that the irreducible mass remains constant for manipulations of the black hole structure deemed `reversible' processes, and cannot decrease under other `irreversible' processes. This means that one can extract the maximum amount of energy (angular momentum or electrostatic energy) from a black hole via a series of reversible processes so that the remaining mass is equal to the irreducible mass at the end of the extraction.

To be precise, the Christodoulou-Ruffini irreducible mass $M_{\mathrm{irr}}$ is given for a Kerr-Newman black hole by the formula
\begin{equation*}
 M^2=\left(M_{\mathrm{irr}}+\frac{Q^2}{4M_{\mathrm{irr}}}\right)^2+\frac{M^2 a^2}{4 M^{2}_{\mathrm{irr}}}.
\end{equation*}
This can be solved for $M_{\mathrm{irr}}$,
\begin{equation*}
 M^{2}_{\mathrm{irr}}=M^{2}-\frac{Q^2}{2}+M\sqrt{M^2-Q^2-a^2}=\frac{r_{+}^2+a^2}{2},
\end{equation*}
where $r_{+}$ is the event horizon, as usual.  In \cite{Christodoulou:1972kt} it was shown that $M_{\mathrm{irr}}$ is constant for reversible processes and can only increase for irreversible processes, so there is a constraint
\begin{equation*}
 \d M_{\mathrm{irr}}\geq 0 ,
\end{equation*}
obeyed by the irreducible mass of the black hole.  

One might think that this constraint could somehow be associated with the second law of thermodynamics, and in fact this turns out to be true.  It was first argued by Bekenstein that black holes must be assigned an entropy \cite{Bekenstein:1973ur, Bekenstein:1974ax}.  The basic idea of the argument is to imagine that an object with some definite entropy falls into a black hole by passing through the event horizon.  The entropy content of the exterior of the black hole is lowered when this object's entropy falls behind the horizon, so the second law of thermodynamics will be violated for an exterior observer unless the black hole itself posesses an entropy which can increase with any addition of mass.

Deriving exactly what this black hole entropy should be is aided by the Penrose-Hawking `area theorem', which states that the area of a black hole's event horizon cannot decrease \cite{Penrose:1971uk, Hawking:1971tu}. Since most processes actually \emph{increase} this area, `[t]his increasing behavior is reminiscent of thermodynamic entropy of closed systems.  Thus it is reasonable that the black hole entropy should be a monotonic function of area, and it turns out to be the simples such function,' to quote Bekenstein himself \cite{Bekenstein:2008}.  

The black hole entropy, $S_{\mathrm{bh}}$, is then given by
\begin{equation*}
 S_{\mathrm{bh}}=\alpha\,A_{\mathrm{hor}}\,,
\end{equation*}
where $\alpha$ is an unknown proportionality factor and $A_{\mathrm{hor}}$ is the event horizon area at a fixed time.  The constant $\alpha$ was determined via the quantum mechanical calculation of Hawking radiation to be \cite{Hawking:1974sw, Parentani:2011}
\begin{equation*}
 \alpha=\frac{1}{4 L^{2}_{\mathrm{p}}}, \qquad L_{\mathrm{p}}=\frac{G \hbar}{c^3},
\end{equation*}
in terms of the Planck length $L_{\mathrm{p}}$.  

Calculating the horizon area,
\begin{equation*}
 A_{\mathrm{hor}}=\int_{0}^{\pi}\d\theta \int_{0}^{2\pi}\d\varphi \sqrt{g_{\theta\theta}g_{\varphi\varphi}}=4\pi(r_{+}^2+a^2),
\end{equation*}
leads finally to an expression for the black hole entropy in terms of the black hole parameters:
\be{bhe1}
 S_{\mathrm{bh}}=4\pi\alpha (r_{+}^2+a^2).
\ee
This gives a relationship between the entropy and the irreducible mass, $S_{\mathrm{bh}}=16\pi\alpha M^{2}_{\mathrm{irr}}$, and the monotonicity of the latter ties it to the area theorem in the form of a second law for black hole entropy:
\begin{equation*}
 \d S_{\mathrm{bh}}=32\pi\alpha M_{\mathrm{irr}}\,\d M_{\mathrm{irr}}\geq 0\,.
\end{equation*}

By varying $(S_{\mathrm{bh}},M,J,Q)$ in \eqref{bhe1}, one obtains a relationship akin to the first law of thermodynamics, namely
\be{bhe2}
\d M=\Omega_{\mathrm{bh}}\,\d J+\Phi_{\mathrm{bh}}\,\d Q+T_{\mathrm{bh}}\,\d S_{\mathrm{bh}}\,,
\ee
where $J=Ma$ is the usual angular momentum and
\begin{equation*}
 \Omega_{\mathrm{bh}}=\frac{a}{r_{+}^2+a^2}, \qquad \Phi_{\mathrm{bh}}=\frac{Q r_{+}}{r_{+}^2+a^2}, \qquad T_{\mathrm{bh}}=\frac{\hbar}{2\pi}\frac{\sqrt{M^2-a^2-Q^2}}{r_{+}^2+a^2}.
\end{equation*}
The quantity $\Omega_{\mathrm{bh}}$ is the angular frequency of a test particle dropped into the black hole just before it enters the horizon, while $\Phi_{\mathrm{bh}}$ is the potential energy of a test particle with electric charge opposite to that of the black hole at the horizon.  

Remarkably, the black hole or Hawking temperature $T_{\mathrm{bh}}$ is the black body radiation temperature of the thermal Hawking radiation emitted from the black hole.  Although classically it can only be determined up to the constant $\alpha$, it is fully determined when quantum effects are taken into account.  Note that when $M$ is large relative to $Q$ and $J$, we have
\begin{equation*}
 T_{\mathrm{bh}}\propto \frac{1}{M},
\end{equation*}
so the larger the mass of a black hole, the lower its temperature.  This temperature is itself often related, in a rather mysterious manner, to the norm of a Killing vector on the horizon, referred to as the \emph{surface gravity}.

These relations can be organized into what is often referred to as the `four laws of black hole thermodynamics' \cite{Bardeen:1973gs}:
\begin{itemize}
 \item[0] \textit{Zeroth Law}: $T_{\mathrm{bh}}$ is uniform over the black hole event horizon.
 \item[1] \textit{First Law}: \eqref{bhe2}, the conservation of energy.
 \item[2] \textit{Second Law}: $\d S_{\mathrm{bh}}\geq0$, entropy of an isolated black hole can never decrease.
 \item[3] \textit{Third Law}: $T_{\mathrm{bh}}$ can never be reduced to zero by a finite number of steps.
\end{itemize}
The status of the third law here is slightly ambiguous: since $T_{\mathrm{bh}}$ is proportional to the surface gravity, and the surface gravity of an extremal black hole vanishes, this law is equivalent to the statement that an extremal black hole cannot be formed in a finite number of steps.

We omit here any discussion of the quantum properties of black holes, which has been the subject of much recent research.

\subsubsection*{\textit{Related solutions}}

The Kerr-Newman metric is just one example of a stationary axisymmetric solution to the Einstein-Maxwell equations, albeit perhaps the most physically relevant one.  An exhaustive overview of the various other solutions in this class can be found in \S 21 of \cite{Stephani:2003}; here we simply mention a few salient examples.  

While the black hole solutions are characterized by three parameters (mass, spin, and electric charge), the most general stationary axisymmetric type D solutions are functions of \emph{seven} parameters \cite{Plebanski:1976gy}.  In addition to $M$, $a$, and $Q$, these metrics are characterized by an acceleration $b$, a magnetic charge $W$, a `NUT parameter' $n$, and a cosmological constant $\Lambda$.  Various well-known type D solutions depend on certain subsectors of these parameters.  For example, the Taub-NUT metric \cite{Newman:1963yy, Misner:1963fr} depends on $\{M,a,n\}$; Demia\'nski's metrics \cite{Demianski:1976} depend on $\{M,a,Q,W\}$; and the generalized Kerr-NUT metrics \cite{Demianski:1966} depend on $\{M,a,n,Q,W\}$.

Since we usually regard the parameters $n$ and $W$ as being unphysical, and a uniform acceleration $b$ hardly ever arises, it is clear that the Kerr-Newman solution occupies the most general section of the physically relevant, asymptotically flat parameter space.  With the addition of a cosmological constant $\Lambda$ of positive (or negative) sign, one obtains the Kerr-Newman-dS (-AdS) black hole \cite{Carter:1968ks, Plebanski:1976gy}, which continues to have spherical event horizon topology, although the precise value of $r_{+}$ now depends on the (anti-)de Sitter space cosmological constant as well as $M$, $a$, and $Q$.  The Kerr-Newman-AdS solution has proved useful in the study of black hole thermodynamics and holography (c.f., \cite{Caldarelli:1999xj, Hartman:2008pb, Chen:2010bh}).

One can also imagine a `multi-Kerr-Newman' configuration, in which space-time contains several charged spinning sources.  Is it possible to find a solution of the field equations describing this situation?  It turns out that in the special case where $|Q|=M$, the Kerr-Newman metric is `conformastationary', a property which defines the metric in terms of solving a potential equation.  When two Kerr-Newman sources, each having this property $|Q_{1,2}|=M_{1,2}$, are arranged with parallel or anti-parallel angular momenta, an explicit metric can be written down \cite{Parker:1973bv, Kobiske:1974mp}.  Unfortunately, as long as $a_{1,2}\neq0$, such metrics will contain naked singularities, so only multi-black hole solutions which are static seem to pass the cosmic censorship test \cite{Hartle:1972ya}.\footnote{In the non-rotating case, extremal Reissner-Nordstr\"om black holes $|Q|=M$ can be embedded in $\cN=2$ supergravity as BPS solutions.  Hence, the existence of multi-black hole solutions in this case can be understood as the no force condition of supersymmetry.} 

\subsubsection*{\textit{Higher dimensions}} 

For the sake of completeness, we briefly touch on higher-dimensional general relativity (and its generalizations), where black holes play an important role (see \cite{Emparan:2008eg, Reall:2012it} for reviews).  It is well-known that the Schwarzschild solution continues to exist in space-times of dimension greater than four, where it is the unique static, spherically symmetric solution to the vacuum Einstein equations \cite{Gibbons:2002bh}.  However, the situation for stationary, asymptotically flat solutions is much more complicated in higher dimensions.  Indeed, perturbative calculations reveal that non-uniqueness (i.e., distinct solutions characterized by the \emph{same} asymptotic conserved quantities) is a generic feature in dimensions higher than four, with explicit examples available in five dimensions.\footnote{It should be noted that there are still uniqueness theorems in higher dimensions, but the data characterizing the black hole is no longer limited to asymptotic charges (c.f., \cite{Hollands:2012xy}).}

The first of these is the Myers-Perry black hole \cite{Myers:1986un}, which can be thought of as the higher-dimensional analogue of the Kerr solution.  This has a spherical event horizon, and is parametrized by its mass and angular momenta (of which there are $\lfloor(d-1)/2\rfloor$ in $d$ space-time dimensions), but even the allowed ranges of these parameters begin to differ dramatically from the four dimensional story.  For instance, in five dimensions, there is an upper bound on the angular momenta, analogous to the extremality bound in $d=4$.  However, for $d>5$ one of the black hole's angular momenta can be arbitrarily large provided the others vanish, although this ultraspinning limit has potential stability issues \cite{Emparan:2003sy}.

There is a second asymptotically flat solution of the vacuum Einstein equations for $d=5$ which is also parametrized by mass and angular momenta: the black ring \cite{Emparan:2001wn}.  This solution has `cylindrical' event horizon topology, with cross-sections $S^1\times S^2$, and can be thought of as describing a loop of matter with tension (roughly, a `massive rubber band') which is rotating sufficiently fast to prevent its collapse due to gravity.  The existence of the black ring demonstrates the failure of the no hair theorem: both a Myers-Perry black hole and a black ring can have the same mass and angular momenta.  Indeed, two distinct black rings can also be described by the same set of parameters.  The black ring and Myers-Perry black hole can be superimposed to give the `Black Saturn' solution \cite{Elvang:2007rd}, which is important as a first example of a well-behaved (i.e., regular, asymptotically flat, stationary, etc.) mutli-black hole solution of Einstein's equations.

Given the existence of the Myers-Perry black hole, it at first seems natural to suppose that a charged version will exist by analogy with four dimensions.  Unfortunately, this does not seem to be the case.  Much like the Kerr metric, the Myers-Perry solution was found by looking for a metric of Kerr-Schild form.  To extend this to a solution of the Einstein-Maxwell equations, one usually looks for a Maxwell field with potential $A_{\mu}=\alpha l_{\mu}$, where $\alpha$ is some function and $l_{\mu}$ is the 1-form defining the Kerr-Schild metric \eqref{KS1}.  Unfortunately, a solution of the field equations for $d>4$ has not been found for a metric and Maxwell field of this Kerr-Schild form.

The issue can be summarized as follows.  In four dimensions, a charged, spinning Kerr-Schild solution has the same principal 1-form as its uncharged counterpart: the Schwarzschild and Reissner-Nordstr\"om metric have $l_{\mu}\d x^{\mu}=\d u$, while Kerr and Kerr-Newman have $l_{\mu}\d x^{\mu}=\d u-a\sin^{2}\theta\d\phi$.  This is simply no longer true in higher dimensions.  Even so, one could also hope to apply the complex transformation algorithm directly to the Myers-Perry solution, but this too has not met with success.  Approximate (i.e., perturbative) solutions have been constructed for `small' charge or angular momenta (c.f., \cite{Caldarelli:2010xz}).

In fact, the best known analogy for the Kerr-Newman black hole in five dimensions is no longer a solution of Einstein-Maxwell theory, but rather $\cN=4$ supergravity (resulting from the compactification of type II or heterotic string theory in ten dimensions).  In this case, the underlying bosonic theory is actually Einstein-Maxwell-Chern-Simons theory (c.f., \cite{Gauntlett:1998fz}).  The stationary, asymptotically flat, charged, and spinning black hole solution in this context is the BMPV black hole \cite{Breckenridge:1996is}.  This solution (and it's non-spinning version, the Strominger-Vafa black hole \cite{Strominger:1996sh}) has played an important role in the study of black hole entropy and thermodynamics. 

Finally, we remark that the concept of algebraic speciality, which has underlined virtually all of this review, changes in higher dimensions.  In four dimensions, the notion of algebraic speciality is captured by degenerate principal null vectors, which are also described in terms of Weyl spinors \cite{Penrose:1984}.  However, for higher dimensions, these vector and spinor approaches lead to \emph{different} definitions of algebraic speciality \cite{DeSmet:2002fv, Coley:2004jv}.  Furthermore, the applicability of these (and other) classifications in finding novel solutions to the field equations has proven rather limited.

\acknowledgments

It is a pleasure to thank Marek Demia\'nski, Pau Figueras, J\"org Frauendiener, and Aidan Keane for various comments and discussions.  The work of TMA is supported by a Title A Research Fellowship at St. John's College, Cambridge.

\bibliography{KN}
\bibliographystyle{JHEP}

\end{document}